\documentclass[aps,prd,twocolumn,showpacs,groupedaddress,floatfix]{revtex4}

\usepackage{graphicx}
\DeclareGraphicsExtensions{.eps, .eps, .jpg, .png}

\def\la{~\mbox{\raisebox{-.6ex}{$\stackrel{<}{\sim}$}}~}
\def\ga{~\mbox{\raisebox{-.6ex}{$\stackrel{>}{\sim}$}}~}

\def\ltap{\ \raise.3ex\hbox{$<$\kern-.75em\lower1ex\hbox{$\sim$}}\ }
\def\gtap{\ \raise.3ex\hbox{$>$\kern-.75em\lower1ex\hbox{$\sim$}}\ }
\def\gl{\ \raise.5ex\hbox{$>$}\kern-.8em\lower.5ex\hbox{$<$}\ }
\def\roughly#1{\raise.3ex\hbox{$#1$\kern-.75em\lower1ex\hbox{$\sim$}}}

\begin{document}



\title{Naturally inflating on steep potentials through electromagnetic dissipation}


\author{Mohamed~M.~Anber and Lorenzo Sorbo}

\affiliation{Department of Physics, University of Massachusetts, Amherst, MA 01003}



\begin{abstract}

In models of natural inflation, the inflaton is an axion-like particle. Unfortunately, axion potentials in UV-complete theories appear to be too steep to drive inflation. We show that, even for a steep potential, natural inflation can occur if the coupling between axion and gauge fields is taken into account. Due to this coupling, quanta of the gauge field are produced by the rolling of the axion. If the coupling is large enough, such a dissipative effect slows down the axion, leading to inflation even for a steep potential. The spectrum of perturbations is quasi-scale invariant, but in the simplest construction its amplitude is larger than $10^{-5}$. We discuss a possible way out of this problem.

\end{abstract}


\maketitle


Axion-like particles are the simplest spin-zero degrees of freedom with a nontrivial radiatively stable potential. Moreover, they are abundant in string theory. As a consequence, axions provide excellent inflaton candidates in a UV complete theory that includes gravity. Inflation driven by axions was proposed as early as in 1990 as {\em natural inflation}~\cite{Freese:1990rb}. The axion potential is radiatively stable thanks to a (broken) shift symmetry, and has the form $V(\Phi)=\Lambda^4[\cos(\Phi/f)+1]$, where $f$ is the axion constant. Neglecting all interactions of $\Phi$ apart from those in $V(\Phi)$ and those with gravity, the condition for inflation is that $V(\Phi)$ is flat in units of the Planck scale (i.e., $|V'|\ll V/M_P$, $|V''|\ll V/M_P^2$) for a sufficiently wide range of $\Phi$. In the case of the axion, these conditions are equivalent to $f\gg M_P$. Unfortunately, string theory appears not to allow such large values of $f$~\cite{Banks:2003sx}. Moreover, $f\lesssim M_P$ appears also as a consequence of the ``gravity as the weakest force'' conjecture of~\cite{ArkaniHamed:2006dz}.

In this paper, we show that natural inflation can be realized also {\em for a steep potential}. Our mechanism relies on the coupling of the inflaton to gauge fields through the operator $\Phi\,F_{\mu\nu}\tilde{F}^{\mu\nu}$. As $\Phi$ rolls down its potential, it provides a time-dependent background for the gauge field whose vacuum fluctuations are thus amplified into physical excitations. This production of quanta of gauge field occurs at the expenses of the kinetic energy of the inflaton, slowing it down. If the coupling between $\Phi$ and $F_{\mu\nu}$ is strong enough, such a dissipation effect can allow to obtain a sufficiently long period of inflation even if $f\ll M_P$. This way, the question of finding natural inflation in string theory is formulated in a new way: is there, among the many axions of string theory~\cite{Arvanitaki:2009fg}, one whose coupling to the gauge fields is large enough? We will argue in Appendix B that the answer to this question might be positive.

\section{Generation of the gauge field}

We consider natural inflation with a pseudoscalar inflaton $\Phi$ coupled to a $U(1)$ gauge field~\footnote{For simplicity we will assume that this is not a Standard Model gauge field. Such a possibility would lead to additional signatures of our scenario that, while out of the scope of the present paper, would be interesting to study.}. The Lagrangian density of the system is given by
\begin{equation}
{\cal L}=-\left[\frac{1}{2}\left(\partial\Phi \right)^2+V(\Phi)+\frac{1}{4}F_{\mu\nu}F^{\mu\nu}+\frac{\alpha}{4f}\,\Phi F_{\mu\nu}\tilde F^{\mu\nu} \right],
\end{equation}
where $V(\Phi)=\Lambda^4\left[1+\cos(\Phi/f) \right]$ with $f\la M_P$~\footnote{Other models of inflation driven by a pseudoscalar (such as~\cite{McAllister:2008hb,Kaloper:2008fb}) with a different form of $V(\Phi)$ have been proposed. The arguments of our paper can extended straightforwardly to these scenarios.}. The parameter $\alpha$ is a dimensionless measure of the coupling of $\Phi$ to the gauge field.

The equations of motion for the system are
\begin{eqnarray}\label{a11}
&&\Phi''+2\,a\,H\,\Phi'-\nabla^{2}\Phi+a^{2}\frac{dV(\Phi)}{d\Phi}=\frac{\alpha}{f}a^{2}\vec E \cdot \vec B \,,\nonumber\\
\label{a12}
&&\vec E'+2\,a\,H\,\vec E-\nabla\times\vec B=-\frac{\alpha}{f}\,\Phi'\,\vec B-\frac{\alpha}{f}\,\vec\nabla\Phi\times\vec E\,,\nonumber\\
\label{a13}
&&\vec\nabla \cdot\vec E=-\frac{\alpha}{f}(\vec\nabla \Phi)\cdot \vec B \mbox{ ,}
\end{eqnarray} 
where $ H\equiv a'(\tau)/a^{2}(\tau)$ and where the prime denotes differentiation with respect to the conformal time $\tau$.
The Bianchi identities read $\vec B'+2\,a\,H\,\vec B+\vec\nabla\times\vec E=0$ and $\vec\nabla\cdot\vec B=0$. Since the inflaton is homogeneous, $\vec\nabla \Phi=0$, we can introduce the vector potential $\vec{A}\left(\tau,\,\vec{x}\right)$, with $a^2\vec{B}=\vec\nabla\times\vec{A}$, $a^2\vec{E}=-\vec{A}'$. Then, the equations for $\vec{A}$ read
\begin{eqnarray}
\label{a15}
\left(\frac{\partial^{2}}{\partial \tau^{2}}-\nabla^{2}-\alpha\,\frac{\Phi'}{f}\,\vec\nabla\times \right)\vec A=0,\,\,\, \vec\nabla\cdot\vec A=0\,.
\end{eqnarray} 

In order to study the generation of the electromagnetic field induced by the rolling pseudoscalar, we promote the classical field $\vec A(\tau,\vec x)$ to an operator $\vec{ \hat{A}}\left(\tau,\,\vec{x}\right)$. We decompose $\vec{\hat A}$ into annihilation and creation operators 
\begin{eqnarray}\label{a16}
\vec{\hat A} =\sum_{\lambda=\pm}\int \frac{d^3k}{\left(2\pi \right)^{3/2}}\left[\vec\epsilon_\lambda(\vec k)\,A_\lambda(\tau,\vec k)\,a_\lambda^{\vec k}\,e^{i\vec k\cdot\vec x}+{\mathrm {h.c.}}\right],
\end{eqnarray}
where the helicity vectors $\vec\epsilon_\pm$ are defined in such a way that $\vec k\cdot \vec \epsilon_\pm=0$, $\vec k\times \vec \epsilon_\pm=\mp i|\vec k|\vec\epsilon_\pm$. Then, $A_\pm$ must satisfy the equations $A_{\pm}''+(k^2 \mp \alpha\,k\,\Phi'/f)A_{\pm}=0$.

Since we are looking for inflationary solutions, we assume $a\left(\tau\right)\simeq -1/(H\,\tau)$, and $d\Phi/dt\equiv \dot\Phi_0=$constant. Hence, the equation for $A_\pm$ reads
\begin{equation}
\label{d3}
\frac{d^{2}A_\pm(\tau,\, k)}{d\tau^{2}}+\left[k^{2}\pm 2\,k\,\frac{\xi}{\tau} \right]A_\pm(\tau,\, k)=0\mbox{ ,}
\end{equation} 
where we have defined 
\begin{equation}
\xi\equiv\alpha\frac{\dot\Phi_0}{2\,f\,H}\,\,,
\end{equation}
We will be interested in the case $\xi\ga{\cal {O}}\left(1\right)$.

Depending on the sign of $\xi$, one of the two solutions $A_{+}$ or $A_{-}$ in (\ref{d3}) will develop an instability. In the following analysis we will assume without loss of generality that $\alpha>0$ and $\dot\Phi>0$ (which implies $V'(\Phi)<0$) so that $\xi>0$.

The solution that reduces to positive frequency for $|\vec k|\tau\rightarrow -\infty$ is $A_\pm(\tau,\, k)=[i\,F_0(\pm\xi,\,- k\,\tau)+G_0(\pm\xi,\,- k\,\tau)]/\sqrt{2\,k}$, where $F_0$ and $G_0$ are the regular and irregular Coulomb wave functions. The mode $A_+$ is rapidly amplified: when the second term in brackets in eq.~(\ref{d3}) dominates over the first one, $\vert k\tau\vert\ll 2\xi$, $A_+$ is approximated by
\begin{equation}
\label{approx}
A_+(\tau,\, k)\simeq 
\frac{1}{\sqrt{2\,k}}\left(\frac{ k}{2\,\xi\,aH}\right)^{1/4}e^{\pi\,\xi-2\,\sqrt{2\xi \,k/aH}}.
\end{equation}
$A_+$ is thus amplified by a factor $e^{\pi\xi}$. On the other hand, the modes $A_-$ are not amplified by the rolling inflaton, and from now on we will ignore them.

\section{The slow roll solution.}

We can now estimate the backreaction of the gauge field on the inflaton, that is described by the term on the right hand side of eq.~(\ref{a11}) (note that the backreaction of the produced gauge field on the inflaton was studied, in a model with different couplings, in~\cite{Watanabe:2009ct}). Using the decomposition of $\vec A$ described above, we get
\begin{eqnarray}\label{edotb}
\langle  \vec E\cdot \vec B \rangle=-\frac{1}{a^4}\int\frac{d^3k}{\left(2\pi\right)^3}\frac{|\vec k|}{2}\frac{\partial}{\partial\tau}\left(\left| A_+\right|^2
-\left| A_-\right|^2\right),
\end{eqnarray} 
that can be calculated by using the approximate expression~(\ref{approx}) and $A_-\simeq 0$. Cutting off the integral at $k_c\simeq 2\,\xi\, H\, a(\tau)$ and approximating $dA_+/d\tau\simeq \sqrt{2\xi\,k\,aH}\,A_+$, we obtain
\begin{equation}\label{edotb}
\langle  \vec E\cdot \vec B \rangle \simeq -\left(\frac{H}{\xi}\right)^4e^{2\pi\xi}\times \left[\frac{1}{2^{21}\pi^2}\int_{0}^{8\xi}dx\,x^7\,e^{-x}\right]\,.
\end{equation} 
Since we assume $\xi\ga 1$, we will send the upper limit of integration in the above equation to infinity, and we denote by ${\cal I}\equiv 7!/(2^{21}\,\pi^{2})\simeq 2.4\times 10^{-4}$ the resulting numerical value of the quantity in brackets.

We then plug $\langle  \vec E\cdot \vec B \rangle$ into eq.~(\ref{a11}) that, in physical time $t$, now reads
\begin{equation}\label{final form in cosmic time}
\frac{d^2\Phi}{dt^2}+3H\frac{d\Phi}{dt}+V'(\Phi) =-\frac{{\cal I}\,\alpha }{f} \left(\frac{H}{\xi}\right)^4e^{2\pi\xi}\,.
\end{equation}
Since we are interested in finding inflationary solutions where slow roll is supported by the dissipation into electromagnetic modes, we assume that both $\ddot \Phi$ and $3\,H\,\dot\Phi$ are negligible with respect to $V'(\Phi)$. In this case, 
an approximate solution of eq.~(\ref{final form in cosmic time}) is 
\begin{equation}
\xi\simeq\frac{1}{2\pi}\log\left[\frac{9}{{\cal I}\,\alpha}\frac{M_P^4\,f\,|V'(\Phi)|}{V^2(\Phi)} \right]\,,
\end{equation}
where we have assumed $3\,M_P^2\,H^2 =\frac{1}{2}\dot\Phi^2+V(\Phi)+\frac{1}{2}(\vec E^2+\vec B^2)\simeq V(\Phi)$ (we will check below the regime of validity of these assumptions). Given the logarithmic dependence on $V(\Phi)$, $\xi$ will never be larger than ${\cal {O}}(10)$. Indeed, unless $\Phi$ is very close to an extremum of $V$ (and with $\alpha$ not exponentially large or small) then $\xi\sim \frac{2}{\pi}\,\log\left[ M_P/\Lambda\right]$. If $\Lambda\sim 10^7$~GeV, so that the reheating temperature is much smaller than $10^8$~GeV (see below for the estimate of the reheating temperature) and overproduction of gravitinos is avoided, then $\xi\simeq 20$. 

We can now explore the part of the parameter space that leads to inflation. Constraints derive from the following requirements:

{\em (i) the Hubble parameter.} We first want to approximate $H^2\simeq V(\Phi)/3\,M_P^2$, the same relation that holds in standard slow roll inflation. This requires that both $\langle \vec E^2+\vec B^2 \rangle$ and $\dot\Phi^2$ be negligible with respect to $V(\Phi)$. With the same technique used to estimate $\langle \vec E\cdot\vec B\rangle$, we obtain
\begin{equation}
\label{rhobtotfin}
\frac{1}{2}\langle \vec E^2+\vec B^2\rangle=\frac{6!\,e^{2\pi\xi}}{2^{19}\,\pi^2}\,\frac{H^4}{\xi^3}\simeq \frac{4}{7}\,\frac{\xi}{\alpha}\,f\,V'(\Phi)\,\,,
\end{equation}
where we have used the slow-roll equation ${\cal I}\,\alpha\, \left(H/\xi\right)^4\,e^{2\pi\xi}\simeq f\,|V'\left(\Phi\right)|$. Eq.~(\ref{rhobtotfin}) shows that for $\alpha\gg \xi$ the energy in the electromagnetic field can be neglected with respect to the energy in the inflaton unless we are close to the bottom of the potential. Indeed, by approximating 
$V(\Phi)\propto \Phi^2$ near its minimum, we see that when 
$\Phi\la \Phi_{\mathrm {RH}}\equiv \xi\,f/\alpha$ the energy in electromagnetic modes cannot be neglected any more. This is the point where reheating begins. The energy density at reheating is $\sim \Lambda^4\,\Phi_{\mathrm {RH}}^2/f^2$, so that the reheating temperature will be of the order of $\Lambda\,\sqrt{\xi/\alpha}$.

Next, let us analyze the condition $\dot\Phi^2/2\ll V$. Using $\dot\Phi=2\,f\,H\,\xi/\alpha$, we obtain
\begin{equation}\label{phidotv}
\frac{\dot\Phi^2}{2\,V(\Phi)}=2\,\frac{\xi^2}{\alpha^2}\,\frac{f^2\,H^2}{V(\Phi)}\simeq \frac{2}{3}\,\frac{\xi^2}{\alpha^2}\,\frac{f^2}{M_P^2}\,\,,
\end{equation}
that shows that the kinetic energy of the inflaton can be neglected with respect to the potential energy for $(\xi/\alpha)\,(f/M_P)\ll 1$. Since $f\la M_P$ by assumption, $\alpha\gg \xi$ will be sufficient to satisfy this condition too;

{\em (ii) acceleration.} In order to make sure that our solution actually corresponds to an inflating Universe, we compute the slow roll parameter $\epsilon\equiv -\dot{H}/H^2$. Using the equations of motion~(\ref{a11}), we obtain
\begin{equation}\label{epsilon}
\epsilon=\frac{1}{2\,M_P^2\,H^2}\,\left[\dot\Phi^2+\frac{2}{3}\,\left(\vec E^2+\vec B^2\right)+\frac{\vec\nabla\cdot(\vec E\times \vec B)}{3\,a\,H}\right],
\end{equation}
where isotropy of the background implies $\langle \vec\nabla\cdot (\vec E\times \vec B )\rangle=0$. By inserting into eq.~(\ref{epsilon}) the values of $\dot\Phi^2$, $\langle \vec E^2+\vec B^2\rangle$ and $H^2$ found above, we derive the expression 
\begin{eqnarray}\label{eps}
\epsilon\simeq \frac{2\,\xi^2}{\alpha^2}\,\frac{f^2}{M_P^2}+\frac{8}{7}\,\frac{\xi}{\alpha}\,\frac{f\,V'(\Phi)}{V(\Phi)}\,\,.
\end{eqnarray}
By comparing the above equation with eq.~(\ref{rhobtotfin}) and (\ref{phidotv}) we see that, as long  as the conditions in {\em (i)} are satisfied, we will have $\epsilon<1$ and the Universe will be inflating;

{\em (iii) neglecting terms in $\ddot\Phi$ and $3\,H\,\dot\Phi$ in eq.~(\ref{final form in cosmic time}).} The following conditions must be satisfied:
\begin{eqnarray}
{\mathrm {(a)}}&&\frac{3\,H\,\dot\Phi}{V'}\sim \frac{\xi}{2\,\alpha}\,\frac{f\,V/V'}{M_P^2}\ll 1\,\,,\label{conda}\nonumber\\
{\mathrm {(b)}}&&\frac{\ddot\Phi}{V'}\sim \frac{2\xi}{3\alpha}\left(-\epsilon\frac{fV/V'}{M_P^2}+\frac{f^2}{\pi M_P^2}\,\frac{VV''/V'{}^2-2}{\alpha}\right)\ll 1.\label{condb}\nonumber
\end{eqnarray}
Since $f\,V/V'={\cal {O}}(f^2)$ and $V\,V''/V'{}^2={\cal {O}}(1)$ unless we are close to an extremum of the potential, then $\alpha\gg \xi\ga 1$ guarantees that both (a) and (b) hold;

{\em (iv) number of efoldings.} The strongest constraint comes by requiring that inflation lasts for long enough. The number of efoldings is given by 
\begin{equation}\label{3rd constraints}
N_e\simeq\int_{\Phi_i}^{\Phi_f}\frac{H\,d\Phi}{\dot{\Phi}}=\frac{\alpha}{2f}\int_{\Phi_i}^{\Phi_f}\frac{d\Phi}{\xi}\simeq \frac{\alpha}{2\,\xi}\,\frac{\Phi_f-\Phi_i}{f}\,.
\end{equation}
Since the range of variation of $\Phi$ is bounded by $|\Phi_f-\Phi_i|\la\pi\,f$, the above equation implies that $\alpha\ga 2\,\xi\,N_e/\pi$. Hence, for $\xi\simeq 20$ we need $\alpha\simeq 600$ to obtain $45$ efoldings of inflation~\footnote{Note that we estimate $\alpha$ by requiring $N_e\simeq 45$ rather than the usual $N_e\simeq 60$ because the choice $\xi\simeq 20$ corresponds to a ``low scale'' inflation at $10^7$~GeV. With $\Lambda\sim 10^{16}$~GeV, i.e. $\xi\simeq 5$, we would require $N_e> 60$, that would translate into the bound $\alpha> 200$.}.

{\em To sum up}, natural inflation with electromagnetic dissipation will last for $N_e$ efoldings if $\alpha\ga 2\, \xi\,N_e/\pi$, where $\xi\sim (2/\pi)\,\log(10\,\alpha^{-1/4}\,M_P/\Lambda)$. There are no limits on the scale of inflation $\Lambda$ (apart from the obvious requirement $\Lambda\ga$~GeV to allow for Big Bang Nucleosynthesis after reheating). Reheating will occur when $\Phi=\Phi_{\mathrm {RH}}$, with 
\begin{equation}
V'(\Phi_{\mathrm {RH}})\simeq \frac{\alpha}{\xi}\,\frac{V(\Phi_{\mathrm {RH}})}{f}\,.
\end{equation}

\section{Perturbations}

Perturbations are usually generated as the quantum fluctuations of the inflaton are amplified by the evolving background. This is different from our scenario, where inhomogeneities in $\Phi$ are sourced classically by those in the electromagnetic field. The situation is analogous to the one studied in~\cite{Green:2009ds,Barnaby:2009mc}, and we will use similar techniques to analyze it. 

The curvature perturbation $\zeta$ on a uniform energy density final hypersurface is related to the perturbation of the number of efoldings by $\zeta=\delta N\equiv N(x)-\bar{N}$, where $\bar{N}$ is the number of efoldings in the homogeneous background.
If we write the perturbed value of the inflaton field as $\Phi=\Phi_{0}(\tau)+\phi(\tau,\,\vec x)$, then $\zeta=H\,\phi/\dot\Phi_0$. In order to compute the power spectrum of $\zeta$, we must therefore compute the correlators of $\phi$. $\phi$ obeys the equation
\begin{equation}
\phi''+2\,a\,H\,\phi'+\left(-\nabla^2+a^2V''\right)\phi=
-\frac{\alpha}{f}\,a^2\,\delta[\vec E\cdot  \vec B],
\end{equation}
where the fluctuation $\delta[\vec E\cdot \vec B](\tau,\vec x)$ receives two contributions. Besides the intrinsic inhomogeneities in $\vec E\cdot\vec B$ (that would be present also for $\phi=0$), a second component comes from the fact that $\langle \vec E\cdot\vec B\rangle$ depends on $\dot\Phi$. As a consequence, if $\Phi$ is replaced by $\Phi+\phi$, then $\langle \vec E\cdot\vec B\rangle$ will go to
$\langle \vec E\cdot\vec B\rangle+\dot\phi\, \partial \langle \vec E\cdot\vec B\rangle /\partial \dot\Phi$. We therefore write
\begin{equation}\label{rhspert}
\delta[\vec E\cdot  \vec B]\simeq [\vec E\cdot\vec B-\langle \vec E\cdot\vec B \rangle]_{\phi=0}+\frac{\partial \langle \vec E\cdot\vec B\rangle}{\partial \dot\Phi}\dot\phi
\end{equation}
We denote the term in square brackets by $\delta_{\vec E\cdot\vec B}\left(\tau,\,\vec x\right)$. The second term is estimated by observing that $\langle \vec E\cdot\vec B\rangle$ depends on $\dot\Phi$ through $\xi$ and that $\partial \langle \vec E\cdot\vec B\rangle /\partial \xi\simeq 2\pi\langle \vec E\cdot\vec B\rangle$. Using the background equation $\alpha\langle \vec E\cdot\vec B\rangle/f\simeq V'$, the second term of the right hand side of eq.~(\ref{rhspert}) can be written as $\pi\,\alpha\,V'\,\dot\phi/(fH)$. The Fourier transform of the perturbation $\phi$ will then obey the (operator) equation
\begin{eqnarray}\label{eqpert}
\nonumber
\phi''(\vec p)&-&\frac{2}{\tau}\,\left(1-\frac{\pi\,\alpha\,V'}{2\,f\,H^2}\right)\,\phi'(\vec p)+\left(p^2+\frac{V''}{H^2\,\tau^2}\right)\,\phi(\vec p)=\\
\label{mastpert}
&-&\frac{\alpha}{f}\,a^2\int\frac{d^3x}{\left(2\pi\right)^{3/2}}\,e^{-i\,\vec p \,\vec x}\,\delta_{\vec E\cdot\vec B}\left(\tau,\,\vec x\right).
\end{eqnarray}
Denoting by $G\left(\tau,\,\tau'\right)$ the retarded Green function associated to the differential operator acting on $\phi$ in the equation above, the two-point function of the inflaton in momentum space reads
\begin{eqnarray}\label{twoptphi}
&&\langle \phi\left(\vec p\right)\phi\left(\vec p\,{}'\right)\rangle=\frac{\alpha^2}{f^2}\int d\tau'\,d\tau'' G\left(\tau,\tau'\right)\,G\left(\tau,\tau''\right)a'{}^2a''{}^2\times\nonumber\\
&&\times\delta\left(\vec p+\vec p\,{}'\right)\int d^3x\,e^{i\,\vec p \,\vec x}\,\langle \delta_{\vec E\cdot\vec B}\left(\tau',\,0\right)\delta_{\vec E\cdot\vec B}\left(\tau'',\,\vec x\right)\rangle,
\end{eqnarray}
where we use the notation $a'\equiv a\left(\tau'\right)$, $a''\equiv a\left(\tau''\right)$.

Therefore, we must compute the two-point correlator of $\delta_{\vec E\cdot\vec B}$, find the Green function associated to the homogeneous part of eq.~(\ref{eqpert}), and compute the integrals in eq.~(\ref{twoptphi}). The details of this derivation are presented in Appendix A. The final result is  
\begin{equation}
\langle\phi\left(\vec p\right) \phi\left(\vec p\,{}'\right)\rangle\simeq 2\times 10^{-6}\frac{\alpha^2\,e^{4\pi\xi}}{\nu_+^2\,f^2}\,\frac{\delta\left(\vec p+\vec p\,{}'\right)}{p^3}\,\frac{H^4}{\xi^8}\left(\frac{2^5\xi p}{a\,H}\right)^{2\nu_-},
\end{equation}
where (see appendix A)
\begin{eqnarray}
&&\nu_+\simeq \pi\alpha V'/(f\,H^2)\propto\alpha\,M_P^2/f^2\gg 1\,, \nonumber\\
&&\nu_-\simeq V'' f/(\pi\alpha V')\propto 1/\alpha\ll 1\,.
\end{eqnarray}

The curvature perturbation is ${\cal P}_\zeta\equiv p^3\,H^2\,\langle \phi\phi\rangle/[2\,\pi^2\,\dot\Phi_0^2\,\delta(\vec p+\vec p\,{}')]$. For generality, we extend the result to the case where the theory contains ${\cal {N}}$ gauge fields. It is straighforward to see that for ${\cal {N}}\neq 1$ the constraints on the parameters found above do not change. However, the different contributions to the two point function of $\delta_{\vec E\cdot\vec B}$ add incoherently, so that ${\cal P}_\zeta$ is suppressed by a factor $1/{\cal {N}}$. Taking this suppression into account, and using $\alpha(H/\xi)^4\,e^{2\pi\xi}=f\,|V'|/{\cal {I}}$, we finally obtain
\begin{equation}
{\cal P}_\zeta\simeq \frac{5\times 10^{-2}}{{\cal {N}}\,\xi^2}\,\left(\frac{2^5\,\xi\, p}{a\,H}\right)^{2\,\nu_-}.
\end{equation}
The spectral index of the scalar perturbations is
\begin{equation}\label{nminusone}
n-1=2\,\nu_-\simeq \frac{2}{\pi\,\alpha}\,\frac{f\,V''(\Phi_0)}{V'(\Phi_0)}\,\,.
\end{equation}
While the sign of $V'$ does not change during inflation, $V''$ crosses zero. As a consequence, the spectrum can be either red or blue depending on the value of $\Phi$ at the time the relevant scale exited the horizon.

The amplitude ${\cal P}_\zeta\sim 0.05/{\cal {N}}\xi^2$ matches the COBE normalization $2.5\times 10^{-9}$ only for large values of ${\cal {N}}$, since $\xi={\cal {O}}(10)$. In particular, if we assume $\xi\simeq 20$, then we need ${\cal N}\simeq 5\times 10^4$ to obtain perturbations with the observed amplitude. Such a large number of gauge fields can be obtained for instance if the theory contains ${\cal N}$ branes, each with its own $U(1)$ gauge field.  Alternatively, one can obtain ${\cal N}$ gauge fields by considering a gauge group $SU(\sqrt{{\cal N}})$ (for instance, ref.~\cite{Rosenhaus:2009cs} considers groups as large as $SU(520)$), that might be obtained on a stack of $\sqrt{{\cal N}}\simeq 200$ branes. This option has the advantage that the different gauge fields have automatically the same coupling $\alpha$ to the inflaton. Note that in the case of nonabelian groups, one should in principle take into account the self interaction of the gauge fields. However, as long as the gauge self-coupling is weak, its effects appear only at higher order, and can be consistently neglected.

\section{Conclusions and future directions}

We have shown that it is possible to realize inflation on a steep axionic potential, provided the inflaton has a sufficiently strong coupling (of the order of $\sim (10^2-10^3)/f$) to a gauge field. Remarkably, this scenario can accommodate inflation at any energy scale and for any value of $f\la M_P$. Unfortunately, the simplest version of the scenario gives an exceedingly large amplitude of scalar perturbations. With a sufficiently large number of gauge fields it is possible to reduce such amplitude to the observed value. It would be interesting to see whether such a suppression could be achieved by other mechanisms.

We still need to show that $\alpha$ as large as $10^2 - 10^3$ can be realized. In Appendix B we show two examples (one directly derived from a string construction, the other based on extra dimensions) where such large values of the coupling can be obtained.

Our mechanism is in spirit similar to that at work in warm~\cite{Berera:1995ie} and trapped inflation~\cite{Kofman:2004yc,Green:2009ds}, that also use dissipation to realize slow roll. The main difference from these models is that our inflaton is coupled to a derivative to the produced field. This allows to achieve a stationary dissipative process without relying on the more complicated structures invoked in those models.

A number of details of this scenario still need to be explored. First, since perturbations are sourced by the inhomogeneities in the gauge field, it is important to study nongaussianities in this models. Future work also involves the generation of gravitational waves, and it will be interesting to study whether a consistency relation similar to that of standard slow roll inflation holds also in this case. One final question concerns parity violation. Since our electromagnetic field is maximally parity violating~\cite{Anber:2006xt}, the gravitational waves produced by the electromagnetic modes could generate a nonvanishing $\langle TB\rangle$ correlation~\cite{Lue:1998mq} in the Cosmic Microwave Background.\\

{\bf \noindent Acknowledgments} We thank Nemanja Kaloper, David Langlois, Anthony Lewis and Alessandro Tomasiello for discussions. This work is partially supported by the NSF grant PHY-0555304. 

\begin{appendix}

\section{Calculation of the power spectrum}

As discussed in section III, the curvature perturbation power spectrum is related the two point function of the perturbations of the scalar $\phi$, eq.~(\ref{twoptphi}). In this appendix we derive the power spectrum ${\cal P}_\zeta$ starting from that expression. In order to perform this calculation, we will first need to compute the two point correlator of $\delta_{\vec E\cdot\vec B}$ and the propagator $G(\tau,\,\tau')$.

\subsection{The two point function of $\delta_{\vec E\cdot\vec B}$}

By using the definitions and the properties of section 3 and assuming $A_-\left(\tau,\,\vec{k}\right)\simeq 0$, the correlator is given by
\begin{widetext}
\begin{eqnarray}
&&\int d^3x\,e^{i\,\vec p \,\vec x}\,\langle 0|\delta_{\vec E\cdot\vec B}\left(\tau',\,0\right)\delta_{\vec E\cdot\vec B}\left(\tau'',\,\vec x\right)|0\rangle=\frac{1}{a\left(\tau'\right)^4\,a\left(\tau''\right)^4}\,\int\frac{d^3 k}{\left(2\pi\right)^3} \left|\vec k\right| \left|\vec\epsilon_+(-\vec k )\cdot \vec\epsilon_+(\vec p+\vec k)\right|^2\times\nonumber\\
&&\times\left\{\left|\vec p+\vec k\right|\,A'_+\left(\tau',\,-\vec k\right)\,A_+\left(\tau',\,\vec p+\vec k\right)\,A'{}^*_+\left(\tau'',\,\vec p+\vec k\right)\,A_+^*\left(\tau'',\,-\vec k\right)+\right.\nonumber\\
&&+\phantom{\{}\left.\left|\vec k\right|\,A'_+\left(\tau',\,\vec p+\vec k\right)\,A_+\left(\tau',\,-\vec k\right)\,A'{}^*_+\left(\tau'',\,\vec p+\vec k\right)\,A_+^*\left(\tau'',\,-\vec k\right)\right\}\,\,.
\end{eqnarray}
\end{widetext}
In principle, the above expression should be renormalized. However, the only part that matters to the generation of inhomogeneities in $\phi$ corresponds to the wavelengths larger than $(2\,\xi\,H)^{-1}$, where the electromagnetic field has large occupation numbers and can be treated as a classical source. In this regime, the function $A_+(\tau,\,\vec{k})$ is well approximated by eq.~(\ref{approx}). As a consequence we can write the above equation as
\begin{eqnarray}\label{int2pf}
&&\frac{e^{4\pi\xi}}{4\,a'{}^4\,a''{}^4}\int\frac{d^3k}{(2\pi)^3}\,\left|\vec\epsilon_+(-\vec k )\cdot \vec\epsilon_+(\vec p+\vec k)\right|^2\times\\
&&e^{-4\sqrt{2\xi/\tilde a H}\,\left(\sqrt{|\vec k|}+\sqrt{|\vec p+\vec k|}\right)}\left\{|\vec k| |\vec p+\vec k|+|\vec k|^{3/2}|\vec p+\vec k|^{1/2}\right\}\,,\nonumber
\end{eqnarray}
where we have defined the function $\tilde a\left(\tau',\,\tau''\right)$ via $2/{\sqrt{\tilde a}}\equiv1/\sqrt{a'}+1/\sqrt{a''}$, with $a'\equiv a\left(\tau'\right)$, $a''\equiv a\left(\tau''\right)$. Note that the integral in the above equation~(\ref{int2pf}) should be cut-off at $|\vec k|,|\vec p+\vec k|\la 2\,\xi\,H\,$Min$\{a',\,a''\}$. However, using a reasoning analogous to the one made after eq.~(\ref{edotb}), we will send the integration limit in $k$ to infinity.

After a change of integration variable, we finally write  the correlator as
\begin{eqnarray}
&&\int d^3x\,e^{i\,\vec p \,\vec x}\,\langle 0|\delta_{\vec E\cdot\vec B}\left(\tau',\,0\right)\delta_{\vec E\cdot\vec B}\left(\tau'',\,\vec x\right)|0\rangle=\nonumber\\
&&=e^{4\pi\xi}\,\frac{\tilde a^5}{a'{}^4\,a''{}^4}\,\frac{H^5}{\xi^5}\,{\cal C}\left(\frac{2^5\,\xi\,\left|\vec p\right|}{\tilde a\,H}\right)\,\,.
\end{eqnarray}
where the function ${\cal {C}}(\kappa)$ (after directing $\vec {p}$ along the $z$ direction) reads
\begin{eqnarray}\label{finalC}
{\cal C}(\kappa)&=&\frac{\kappa^5}{2^{30}\,\pi^3}\,\int d^3q\,\left|\vec\epsilon_+(-\vec q )\cdot \vec\epsilon_+(\hat z+\vec q)\right|^2\times\\
&&\times e^{-\sqrt{\kappa}\,\left(\sqrt{|\vec q|}-\sqrt{|\hat z+\vec q|}\right)}\left|\vec q\right|\,\left|\hat z+\vec q\right|\left\{1+\frac{\left|\vec q\right|^{1/2}}{\left|\hat z+\vec q\right|^{1/2}}\right\}\,,\nonumber
\end{eqnarray}
where $\hat z$ is the unit vector in the $z$ direction.

\subsection{The Green function}

In principle, the Green function for equation~(\ref{mastpert}) can be computed exactly (after assuming $\dot\Phi_0\simeq$constant). We can however get simpler analytical expressions by limiting ourselves to the case of the cosine  potential~$V(\Phi)\propto 1+\cos(\Phi/f)$. In this case, $V'\left(\Phi_0\right)\sim V(\Phi_0)/f$,  $V''\left(\Phi_0\right)\sim V(\Phi_0)/f^2$, $H^2\sim V(\Phi_0)/M_P^2$ and $\alpha\gg 1$ while $f\la M_P$. This allows to see that the coefficient of $d\phi/d\tau$ in eq.~(\ref{mastpert}) is much larger than one. Moreover, we can neglect the term $p^2$ in the coefficient of $\phi$. This is possible for $|p|\ll a\,\sqrt{|V''|}\simeq (M_P/f)\,aH$, a condition that, as long as $f\la M_P/\xi$, is always satisfied if $p\ll 2\xi\,aH$, that is the necessary condition for the validity of eq.~(\ref{approx}).

Once we have taken into account these two approximations, we can find the appropriate Green function, obtained by solving
\begin{eqnarray}
\nonumber
\frac{\partial^2G(\tau,\,\tau')}{\partial \tau^2}&-&\frac{1}{\tau}\,\frac{\pi\,\alpha\,V'\left(\Phi_0\right)}{f\,H^2}\,\frac{\partial G(\tau,\,\tau')}{\partial\tau}\\
&+&\frac{V''\left(\Phi_0\right)}{H^2\,\tau^2}\,G(\tau,\,\tau')=\delta\left(\tau-\tau'\right)\,.
\end{eqnarray}
with $G(\tau',\,\tau')=0,\,(\partial G/\partial \tau)(\tau',\,\tau')=1$. The solution is
\begin{eqnarray}
&&G(\tau,\,\tau')=\left\{
\begin{array}{cl}
\frac{\tau'}{\nu_+-\nu_-}\,\left[\left(\frac{\tau}{\tau'}\right)^{\nu_+}-\left(\frac{\tau}{\tau'}\right)^{\nu_-}\right]\,\,, & \tau>\tau'\\
0\,\,, & \tau<\tau'
\end{array}
\right.\\
\nonumber
&&\nu_\pm\simeq \frac{\pi\,\alpha\,V'\left(\Phi_0\right)}{2\,f\,H^2}\left[1\pm\sqrt{1-\frac{4}{\pi^2}\,\frac{1}{\alpha^2}\frac{V''\left(\Phi_0\right)\,H^2\,f^2}{V'(\Phi_0)^2}}\right]\,\,.\\
\label{nupm}
\end{eqnarray}
The second term under the square root in eq.~(\ref{nupm}) scales as $(f/\alpha\,M_P)^2$ and is much smaller than one. As a consequence we have $\nu_+\simeq \pi\alpha V'/(f\,H^2)\propto\alpha\,M_P^2/f^2\gg 1$ whereas $\nu_-\simeq V'' f/(\pi\alpha V')\propto 1/\alpha\ll 1$.

\subsection{The spectrum}

We are interested in the spectrum at $p\ll aH$. As a consequence, we can neglect the term $(\tau/\tau')^{\nu_+}$ in the expression of the Green function, that goes rapidly to zero. We then collect the results of the above subsections, we use $a=-1/H\tau$ and define the new integration variables $w'=-\left(2^5\,\xi\,\left|\vec p\right|\,\tau'\right)^{-1}$,  $w''=-\left(2^5\,\xi\,\left|\vec p\right|\,\tau''\right)^{-1}$. Therefore, the two point function reads
\begin{eqnarray}
\nonumber
&&\langle \phi\left(\vec p\right) \phi\left(\vec p\,{}'\right)\rangle=\frac{\delta^{(3)}\left(\vec p+\vec p\,{}'\right)}{p^3}\,\frac{{\cal {N}}\,\alpha^2\,H^4}{\nu_+^2\,f^2\,\xi^8}\,\frac{e^{4\pi\xi}}{2^{15}}\,\left(\frac{2^5\,\xi\, p}{a\,H}\right)^{2\nu_-}\\
\nonumber
&&\times\int_{0}^{\frac{aH}{2^5\xi\, p}} dw'\,w'{}^{\nu_--5}\int_{0}^{\frac{aH}{2^5\xi\,p}} dw''\,w''{}^{\nu_--5}\,\tilde w^5\,{\cal C}\left(\tilde w^{-1}\right)\,,\\
\end{eqnarray}
where we have defined $2/\sqrt{\tilde w}\equiv 1/\sqrt{w'}+1/\sqrt{w''}$. We see now that, as long as $\nu_-\ll 1$ the spectrum of perturbations in the inflaton is quasi-scale invariant. To find the normalization, we send $p\rightarrow 0$ in the limits of integration. 

The integral can then be computed by defining the new integration variables $x'=w'{}^{-1/4}$, $x''=w''{}^{-1/4}$ and going to "polar coordinates" $x'=\rho\cos\theta$, $x''=\rho\sin\theta$. We set $\nu_-=0$ to simplify the resulting expressions. We use the expression~(\ref{finalC}) for the function ${\cal C}$. The integrals in $\theta$ and $\rho$ can be now computed explicitly
\begin{eqnarray}
&&\int_{0}^\infty \frac{dw'}{w'{}^{5}}\int_{0}^\infty \frac{dw''}{w''{}^{5}}\,\tilde w^5\,{\cal C}\left(\tilde w^{-1}\right)=\frac{\Gamma(8)\,\Gamma(16)}{2^{27}\,\pi^{\frac{5}{2}}\,\Gamma(\frac{17}{2})}\times\\
&&
\int d^3q\,\left|\vec\epsilon_+(-\vec q )\cdot \vec\epsilon_+(\hat z+\vec q)\right|^2\frac{\left|\vec q\right|\,\left|\hat z+\vec q\right|+\left|\vec q\right|^{3/2}\,\left|\hat z+\vec q\right|^{1/2}}{\left(\sqrt{|\vec q|}+\sqrt{|\hat z+\vec q|}\right)^{16}}\,,\nonumber
\end{eqnarray}
where the integral in $d^3q$ can be evaluated numerically to $\simeq 3.5\times 10^{-4}$.

The final result is
\begin{equation}
\langle\phi\left(\vec p\right) \phi\left(\vec p\,{}'\right)\rangle=\gamma\,\frac{\delta^{(3)}\left(\vec p+\vec p\,{}'\right)}{p^3}\,\frac{{\cal N}\,\alpha^2}{\nu_+^2\,f^2}\,e^{4\pi\xi}\,\frac{H^4}{\xi^8}\,\left(\frac{2^5\,\xi\, p}{a\,H}\right)^{2\nu_-}
\end{equation}
where the numerical factor $\gamma\simeq 2.1\times 10^{-6}$

We can finally find the curvature perturbation ${\cal P}_\zeta=p^3\,H^2\,\langle \phi\phi\rangle/[2\,\pi^2\,\dot\Phi_0^2\,\delta^{(3)}(\vec p+\vec p\,{}')]$. Using $\alpha(H/\xi)^4\,e^{2\pi\xi}=f\,V'/{\cal {I}}$, we obtain
\begin{equation}
{\cal P}_\zeta=\frac{\gamma}{8\,\pi^4\,{\cal {I}}^2}\,\frac{1}{\xi^2}\,\left(\frac{2^5\,\xi\, p}{a\,H}\right)^{2\nu_-}\simeq \frac{5\times 10^{-2}}{\xi^2}\,\left(\frac{2^5\,\xi\, p}{a\,H}\right)^{2\nu_-}.
\end{equation}

If the theory contains ${\cal {N}}$ gauge fields, then the different contributions to the two point function of $\delta_{\vec E\cdot\vec B}$ will sum incoherently, leading to a suppression by a factor of ${\cal {N}}$ of ${\cal P}_\zeta$, i.e., 
\begin{equation}
{\cal P}_\zeta\simeq \frac{5\times 10^{-2}}{{\cal {N}}\,\xi^2}\,\left(\frac{2^5\,\xi\, p}{a\,H}\right)^{2\nu_-}.
\end{equation}

\section{Obtaining $\alpha={\cal {O}}(100)$}

We present here two examples where one can obtain a value of $\alpha$ large enough to allow our mechanism to work.

\subsection{Two axions, two gauge fields}

The first example we consider is based on a lagrangian presented already in~\cite{Choi:1985bz}, that studied axion phenomenology in the dimensionally reduced $E_8\times E_8$ superstring. The low energy theory contains a model independent axion $\Phi_1$ and a model dependent one, $\Phi_2$
\begin{eqnarray}
&&{\cal {L}}=-\frac{1}{2}\left(\partial \Phi_1\right)^2-\frac{1}{2}\left(\partial \Phi_2\right)^2+\\
&&-\frac{1}{4}\left(\frac{\Phi_1}{f_1}+\frac{\Phi_2}{f_2}\right)\,F^a_{\mu\nu}\tilde{F}^a{}^{\mu\nu}-\frac{1}{4}\left(\frac{\Phi_1}{f_1}-\frac{\Phi_2}{f_2}\right)\,G^a_{\mu\nu}\tilde{G}^a{}^{\mu\nu},\nonumber
\end{eqnarray}
where $F^a_{\mu\nu}$ and $G_{\mu\nu}^a$ denote the field strength of two different gauge groups.

We then redefine 
\begin{eqnarray}
\Phi_1&=&\frac{f_2}{\sqrt{f_1^2+f_2^2}}\,\Phi-\frac{f_2}{\sqrt{f_1^2+f_2^2}}\,\tilde\Phi\,\,,\nonumber\\
\Phi_2&=&\frac{f_1}{\sqrt{f_1^2+f_2^2}}\,\Phi+\frac{f_1}{\sqrt{f_1^2+f_2^2}}\,\tilde\Phi\,\,,
\end{eqnarray}
so that the fields $\Phi$ and $\tilde\Phi$ are canonically normalized and the interactions between the axions and the $F_{\mu\nu}$ and $G_{\mu\nu}$ read
\begin{eqnarray}
{\cal {L}}_{\mathrm {int}}=&&-\frac{\Phi}{4\,f_{\mathrm {eff}}}\left(G^a_{\mu\nu}\tilde{G}^a{}^{\mu\nu}+\frac{f_2^2+f_1^2}{f_2^2-f_1^2}\,F^a_{\mu\nu}\tilde{F}^a{}^{\mu\nu}\right)+\nonumber\\
&&+\frac{\tilde\Phi}{2\,\sqrt{f_1^2+f_2^2}}\,G^a_{\mu\nu}\tilde{G}^a{}^{\mu\nu}\,.
\end{eqnarray}

The field $\tilde\Phi$ can be ignored, and if the $G_{\mu\nu}$ sector gets strongly coupled at energy $\Lambda$, then a potential $\sim \Lambda^4\,\cos(\Phi/f_{\mathrm {eff}})$ is generated. By identifying $\left(f_2^2+f_1^2\right)/\left({f_2^2-f_1^2}\right)=\alpha$ we can obtain $\alpha\gg 1$ by fine-tuning  $f_1\simeq f_2\left(1-1/\alpha\right)$.

\subsection{Extra dimensions}

A second possibility involves extra dimensions. To fix ideas, let us consider a flat, five dimensional bulk compactified with radius $R$. A single Peccei-Quinn field $\eta=v^{3/2}\,e^{i\,\theta}$ lives in the bulk, where we have already fixed the absolute value of $\eta$ to $v^{3/2}$ and we will neglect $\eta$'s radial mode. Assuming that the kinetic term of $\eta$ is canonically normalized, we have that the lagrangian for $a$ reads $-\int d^4x\,dy\,v^3\,(\partial \theta)^2$. Let us now consider two gauge fields $A^1_\mu$ and $A^2_\mu$, both living in the bulk. Both gauge fields will have an axionic coupling to $\theta$, but while the coupling to $A^1_\mu$ will be in the bulk, the coupling to $A_\mu^2$ will be localized on a brane at $y=0$. Therefore, the relevant part of the lagrangian reads
\begin{eqnarray}
\int d^4x\int_{-R/2}^{R/2}&dy&\left[\left(-v^3\left(\partial\theta\right)^2-\frac{1}{4}F^1_{\mu\nu}F^1{}^{\mu\nu}-\frac{1}{4}F^2_{\mu\nu}F^2{}^{\mu\nu}\right.\right.+\nonumber\\
&+&\left.\left.\theta\,\tilde{F}^1_{\mu\nu}\,F^1{}^{\mu\nu}\right)+\frac{\delta(y)}{M}\,\theta\,\tilde{F}^2_{\mu\nu}\,F^2{}^{\mu\nu}\right]\,\,,
\end{eqnarray}
where $M$ is some mass scale that we assume to be of the order of the cutoff scale of the theory. We obtain a four dimensional effective action by integrating on the extra dimensions, and canonically normalizing both the axion and the gauge fields. The resulting lagrangian for the zero modes reads
\begin{eqnarray}
\int d^4x&&\left[\left(\partial\theta\right)^2-\frac{1}{4}F^1_{\mu\nu}\,F^1{}^{\mu\nu}-\frac{1}{4}F^2_{\mu\nu}\,F^2{}^{\mu\nu}+\right.\\
&&+\left.\frac{\theta}{\sqrt{v^3\,R}}\,\tilde{F}^1_{\mu\nu}\,F^1{}^{\mu\nu}+\frac{\theta}{(M\,R)\sqrt{v^3\,R}}\tilde{F}^2_{\mu\nu}\,F^2{}^{\mu\nu}\right]\,\,,\nonumber
\end{eqnarray}

Let us now assume that the second gauge group condenses at some scale $\Lambda$. Then a potential term $\sim \Lambda^4\,\cos(\theta/f)$ will be generated, with $f\simeq M\,R\,\sqrt{v^3\,R}$. The coupling of $\theta$ to the first gauge group will thus have the form $M\,R\,(\theta/f)\,\tilde{F}^1_{\mu\nu}\,F^1{}^{\mu\nu}$ so that $\alpha\simeq M\,R$ can be of the order of $10^2$ provided the extra dimension is large enough.

\end{appendix}

\end{document}